\documentstyle[preprint,eqsecnum,aps,epsfig]{revtex}

\begin{document}
\draft
\preprint{ }
\title{Primary proton spectrum between 200 TeV and 1000 TeV observed with the Tibet
burst detector and air shower array}

\author{
 The Tibet AS${\bf \gamma}$ Collaboration\\
 M.~Amenomori$^{1}$, S.~Ayabe$^{2}$,   Caidong$^{3}$,  Danzengluobu$^{3}$,
 L.K.~Ding$^{4}$,   Z.Y.~Feng$^{5}$,   Y.~Fu$^{6}$,    H.W.~Guo$^{3}$, 
 M.~He$^{6}$,  K.~Hibino$^{7}$,   N.~Hotta$^{8}$,    Q.~Huang$^{5}$,
 A.X.~Huo$^{4}$,  K.~Izu$^{9}$,  H.Y.~Jia$^{5}$,    F.~Kajino$^{10}$, 
 K.~Kasahara$^{11}$,   Y.~Katayose$^{12}$,  Labaciren$^{3}$, J.Y.~Li$^{6}$,
 H.~Lu$^{4}$,   S.L.~Lu$^{4}$,  G.X.~Luo$^{4}$, X.R.~Meng$^{3}$, 
 K.~Mizutani$^{2}$,  J.~Mu$^{13}$,  H.~Nanjo$^{1}$, M.~Nishizawa$^{14}$,
 M.~Ohnishi$^{9}$, I.~Ohta$^{8}$,  T.~Ouchi$^{7}$,  Z.R.~Peng$^{4}$,       
  J.R.~Ren$^{4}$,  T.~Saito$^{15}$,  M.~Sakata$^{10}$, T.~Sasaki$^{10}$, 
  Z.Z.~Shi$^{4}$, M.~Shibata$^{12}$,   A.~Shiomi$^{9}$,  T.~Shirai$^{7}$,
  H.~Sugimoto$^{16}$,  K.~Taira$^{16}$,  Y.H.~Tan$^{4}$,
  N.~Tateyama$^{7}$,    S.~Torii$^{7}$,  T.~Utsugi$^{2}$,  C.R.~Wang$^{6}$,
  H.~Wang$^{4}$,   X.W.~Xu$^{4,9}$,       Y.~Yamamoto$^{10}$,  G.C.~Yu$^{3}$, 
  A.F.~Yuan$^{3}$, T.~Yuda$^{9,17}$,    C.S.~Zhang$^{4}$,    H.M.~Zhang$^{4}$, 
  J.L.~Zhang$^{4}$,   N.J.~Zhang$^{6}$,  X.Y.~Zhang$^{6}$, Zhaxiciren$^{3}$,
  and  Zhaxisangzhu$^{3}$
}
\address{
$^{1}$Department of Physics, Hirosaki University, Hirosaki 036-8561, Japan\\
$^{2}$Department of Physics, Saitama University, Urawa 338-8570, Japan\\
$^{3}$Department of Mathematics and Physics, Tibet University, Lhasa 850000, China\\
$^{4}$Laboratory of Cosmic Ray and High Energy Astrophysics,\\
      Institute of High Energy 
          Physics, Academia Sinica, Beijing 100039, China\\
$^{5}$ Department of Physics, South West Jiaotong University, Chengdu 610031, China\\
$^{6}$ Department of Physics, Shangdong University, Jinan 250100, China\\
$^{7}$ Faculty of Engineering, Kanagawa University, Yokohama 221-8686, Japan\\
$^{8}$ Faculty of Education, Utsunomiya University, Utsunomiya 321-8505, Japan\\
$^{9}$ Institute for Cosmic Ray Research, University of Tokyo, Kashiwa 277-8582, Japan\\
$^{10}$ Department of Physics, Konan University, Kobe 658-8501, Japan\\
$^{11}$ Faculty of Systems Engineering, Shibaura Institute of Technology, 
Omiya 330-8570, Japan\\
$^{12}$ Faculty of Engineering, Yokohama National University, Yokohama 240-0067, Japan\\
$^{13}$ Department of Physics, Yunnan University, Kunming 650091, China\\
$^{14}$ National Institute of Informations, Tokyo 101-8430, Japan\\
$^{15}$ Tokyo Metropolitan College of Aeronautical Engineering, Tokyo 116-0003, Japan\\
$^{16}$ Shonan Institute of Technology, Fujisawa 251-8511, Japan\\
$^{17}$ Solar-Terrestrial Environment Laboratory, Nagoya University, 
Nagoya 464-8601, Japan 
 }

\date{\today}
\maketitle
\begin{abstract}             
Since 1996, a hybrid experiment consisting of the emulsion chamber
and burst detector array and the Tibet-II air-shower array
 has been operated at Yangbajing (4300~m above sea level, 606 g/cm$^2$) in Tibet.
This experiment can detect
air-shower cores, called as burst events, accompanied by
air showers in excess of about 100~TeV. We observed about 4300 
burst events accompanied by air showers during  690 days of
operation and selected 820 proton-induced events with its primary energy
above 200 TeV using a neural network method.
Using this data set, we obtained the energy spectrum of primary
protons in the energy range from 200 to 1000~TeV. The
differential energy spectrum obtained in this energy region
 can be fitted by a
power law with the index of -2.97  $\pm$ 0.06, which is
steeper than that obtained by direct measurements
at lower energies. 
We also obtained  the energy spectrum of helium nuclei at particle energies
around 1000~TeV.
\end{abstract}

\pacs{PACS numbers : 98.70Sa, 95.85Ry, 96.40De,  96.40Pq}

\section{ INTRODUCTION}
Shock acceleration at supernova blast waves gives a good explanation of
the origin of the bulk of  cosmic rays.  It may be well accepted 
that cosmic rays below about 10 TeV are predominantly due to the  explosion
of stars (supernova explosion) into the normal interstellar medium, while
particle acceleration at supernova remnants (SNR's) has an upper limit
of about 100~TeV \cite{1,2}.  Also, there is an argument that the cosmic rays 
from near 10~TeV to several times 1000~TeV very likely  originate in  
the explosion of massive stars into their former stellar wind \cite{3}.
These processes  have been examined to be able to explain the
cosmic ray spectra fairly well up to the highest energy where abundances are known
\cite{4}.  For energies beyond about 1000~TeV, however, there is no consensus.
On the other hand, ground-based
air-shower experiments observe cosmic rays with energies
up to $\sim$100~EeV (10$^{20}$~eV). 
 Measurements so far reported \cite{5} suggest that the  slope of the 
all-particle spectrum in the energy range of about 100 - 1000~TeV 
is somewhat flatter than that observed at lower energies, while at higher energies over
several times 1000~TeV the energy spectrum  becomes steeper with the slope of about
 -3.0.  The break in the overall spectrum at around 1000~TeV 
is often referred to as the ``knee'' in the spectrum.

Clearly, the knee of the primary cosmic ray
spectrum has its origin in  the acceleration and propagation
of high-energy cosmic rays in our Galaxy. The acceleration model by supernova blast waves
 leads to the formation of a  power-law spectrum of particle energies with the
 index of about -2 at sources \cite{1}, and plausible propagation models of their confinement by 
galactic magnetic fields and of their eventual escape from our Galaxy
can explain well a steeper power-law spectrum than that
at the source region \cite{5}, suggesting a rigidity-dependent bending for 
different cosmic ray composition. 
Within the framework of this picture the average mass of primary cosmic
rays before the knee should increase with increasing primary energy.
 In other words, the knee composition 
becomes heavy dominant as the proton spectrum may first bend at an energy of  
about 100~TeV, corresponding to a maximum energy gained by shock acceleration at SNR's \cite{1}.

While the  origin of cosmic rays with energies beyond the knee is still in controversy,
observations of cosmic rays in such a high energy region  may 
naturally stand in need of
 other acceleration mechanisms \cite{6,7}  or new cosmic 
ray sources \cite{8,9}. Among those, one of the most promising  models
 may be  that the cosmic rays come
 from extra-galactic sources such as active galactic nuclei \cite{9},
though the evidence is far from convincing. However,
 such an extra-galactic source
model should  predict proton-enriched primary composition around and beyond the knee. 

Thus,  measurements of the primary cosmic rays around the knee
are  very important and its composition is fundamentally input for understanding the 
particle  acceleration mechanism that pushes cosmic rays  to very high energies.
Among various primary particles,  protons hold the key to the situation and its spectrum
 provides major constraints on the model parameters  of the origin of high-energy cosmic rays.
Because of extremely low and steeply decreasing flux at high energies, however, 
direct measurements of primary 
proton spectrum on board balloons are still limited in the energy region 
lower than a few hundred TeV.  In a recent report by the JACEE 
group \cite{10} it was concluded that the proton spectrum as well as the helium 
spectrum are consistent with power laws with no spectral breaks,
meaning that there is no bending up to the highest energy
they measured (about 800~TeV). However, this is a surmise based on 
statistically sparse  data, so more studies
are required.  On the other hand, most studies on the cosmic ray
composition around the knee have been carried out  with  ground-based
instruments that can observe the various air-shower parameters.
Recently, for example, measurements of muon
content in each air-shower \cite{11} or muons in the deep underground \cite{12,13}, 
measurements of the lateral distribution of air shower Cherenkov lights \cite{14} or 
the maximum depth of shower development using air Cherenkov telescopes \cite{15},  
and  multiparameter measurements of air showers \cite{16}
have been carried out. 
However, the  results obtained by these methods
have been  derived by  indirect ways that may  strongly rely  on how 
 the observed quantities depend on the composition, on the
precision of the measurements,  and on the air-shower and detector 
simulations as well.  Therefore, the  conclusions  
sometimes differ with experiments considerably.

Within the ground-based experiments those which set up at higher altitudes
are preferable. The reasons include, first, the observation level is
close to the maximum of the shower developments induced by cosmic rays with
energies around the knee, so that the energy
determination is more precise and less dependent upon the unknown
composition \cite{17} ; second, the higher energy flux 
in the core region of air
showers can be observed with emulsion chambers or burst detectors \cite{18}. 
High-energy air-shower cores
are sensitive to the intensity of protons in the primary
cosmic rays and also to the composition around the knee.  

A hybrid-experiment of
emulsion chamber and air-shower array at high altitude has a great advantage for studying
 the composition of primary particles at the knee energy
region \cite{19,20}.
In a previous paper \cite{21} we have developed a method to study the
primary cosmic ray composition with a hybrid detector of the emulsion chamber 
and air-shower 
array based on a Monte Carlo simulation. It is shown there that 
an artificial neural network (ANN) can be used as a classifier 
for the species of primary particles since high-energy air-shower cores
accompanying air showers are characterized by several parameters
and that such a hybrid experiment is powerful enough to select the
events induced by protons in the knee energy region.
We have applied a three layered feed forward neural network with
a back-propagation learning algorithm  to the data  obtained with the Tibet burst detector and
the air-shower array \cite{22}.

Here, we report our study on the primary proton spectrum
using the data obtained with the Tibet burst detector and
air-shower array. The experiment, including
the apparatus, its performance and data selection, is 
described  in Sec. II.  Air-shower simulations to compare with the
experimental data are described in 
Sec. III.  The ANN used is briefly introduced 
in Sec. IV. Section V is devoted to the results and discussions
and a brief summary is given in
Sec. VI.

\section{EXPERIMENT}
\subsection{Apparatus}

We started a hybrid experiment of the emulsion chamber, the burst detector 
and the air-shower array  (Tibet-II) at Yangbajing (4300 m above sea 
level, 606 g/cm$^2$), Tibet in 1996 \cite{22}.  The Tibet-II array
consists of 221 scintillation counters of 0.5 m$^2$ each of which are placed on
a 15 m square grid, and which has been
operated since 1995.  Any fourfold coincidence in the detectors 
is used as the trigger 
condition for air-shower events. Under this condition the trigger rate
is about 200 Hz with a dead time of about 12\%  for data taking. The energy 
threshold is estimated 
to be about 7 TeV for proton-induced showers. The precision of the
shower direction determination is about $1^\circ$, which  has been  confirmed
by observing the Moon's shadow \cite{23}. The main aim of Tibet-II is
to search for   gamma ray point sources at energies around 10 TeV. But it 
can also be used for
the measurement of the all-particle  spectrum of  cosmic rays \cite{9}, and for
the study of topics in the knee region by providing information
on the shower size, direction, core position, and arrival time
of each air-shower event to the core detectors \cite{22,24}.

\begin{figure}
\begin{center}

  \vspace*{2cm}

  \epsfig{file=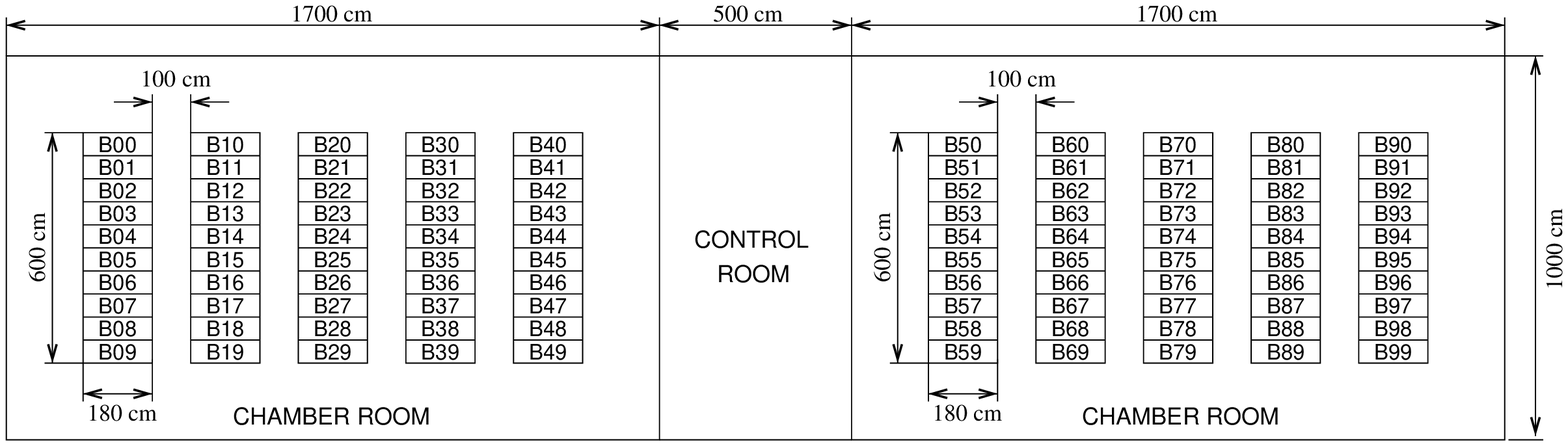,height=4cm} \par  \vspace{1cm}

\caption{ Arrangement of 100 burst detectors set up in two rooms. The area of each burst 
detector is
50~cm $\times$ 160~cm and four emulsion chambers are set up on each burst detector. 
\label{Fig1}}
\end{center}
\end{figure}

\begin{figure}
\begin{center}

  \epsfig{file=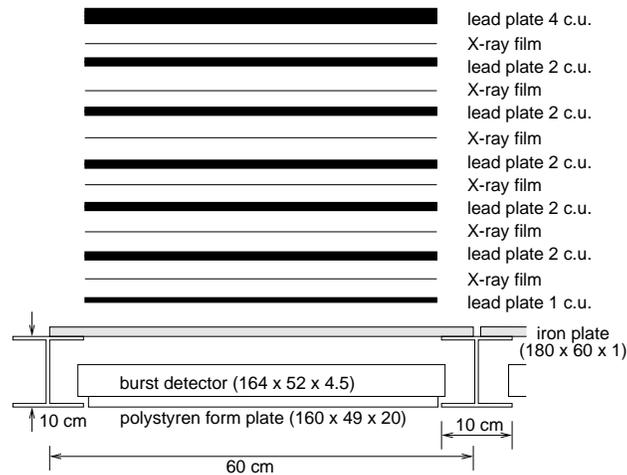,height=8 cm} \par  \vspace{1cm}

\caption{ Schematic side view of each unit of emulsion chamber and burst detector. High sensitive
x-ray films are inserted at every 2 c.u. in emulsion chamber. Total thickness of lead plates is
15~c.u. (7.5~cm).
\label{Fig2}}

\end{center}
\end{figure}

 The emulsion chambers and the burst detectors are 
used to detect high-energy 
air-shower cores accompanied by air showers induced by primary cosmic rays with 
energies above 10$^{14}$ eV. They are separately set up in two rooms as shown
 in Fig. \ref{Fig1}
and placed near the center of the Tibet-II array. 
A basic structure of each  emulsion chamber used here is a multilayered
sandwich of lead plates and photosensitive
x-ray films \cite{18}. Photosensitive layers are set every 2 cascade units (c.u.) 
(here, 1 c.u. is taken to be 0.5 cm) of lead in the
chamber as shown in Fig. \ref{Fig2}.
 There are 400 units of emulsion chamber, each with an area of
40 cm $\times$  50 cm with the total thickness of 15 c.u., giving the total 
sensitive area of 80 m$^2$, and 100
units of burst detectors  each with an effective
area of 160 cm $\times$ 50 cm. Four units of the emulsion chamber are set above
one unit of the burst detector. A 1 cm iron plate is set between the emulsion
chambers and burst detectors.

Each burst detector consists of a plastic scintillator with the size of
160cm $\times$ 50cm and thickness of 2 cm, and four photodiodes (PD's) are
 attached at four
corners of each scintillator to read light signals generated by shower
particles produced in the lead and iron absorber above the
detector. Using the analog-to-digital converter (ADC)
 values from four PD's the total number
(i.e., burst size, $N_b$, for each burst detector) and the position of the 
number-weighted center of all
shower particles that hit a burst detector can be estimated. The 
response of the burst detector is calibrated using electron beams from an 
accelerator and cosmic ray muons. The performance of the burst 
detector and the calibration using
the electron beams are briefly summarized in Appendix A. It is confirmed that
the burst size capable of measuring with each detector ranges from $10^4$ to
$3\times10^6$, roughly corresponding to showers with energies ranging 
from $\sim$~2 to $\sim$~300 TeV. 

A burst event is triggered
when any twofold coincidence of signals from four PD's of a burst
detector appears. Using the 
burst detector array  shown in Fig. \ref{Fig1}, the electromagnetic
 components in the air-shower cores can be measured in the area within 
a radius of several meters.
The coincidence of a burst event and an air-shower event
 is made by their arrival times, and the coincidence of a burst event and
a family event observed in the emulsion chamber is made by their positions
and directions (A burst event and its accompanying air-shower have
the same direction.).

In the following  analysis we  use only the data obtained from all burst detectors and
the Tibet-II array, while the emulsion chamber data will be
reported elsewhere in the  near future.

\subsection{Data analysis}

\begin{figure}
\begin{center}

  \epsfig{file=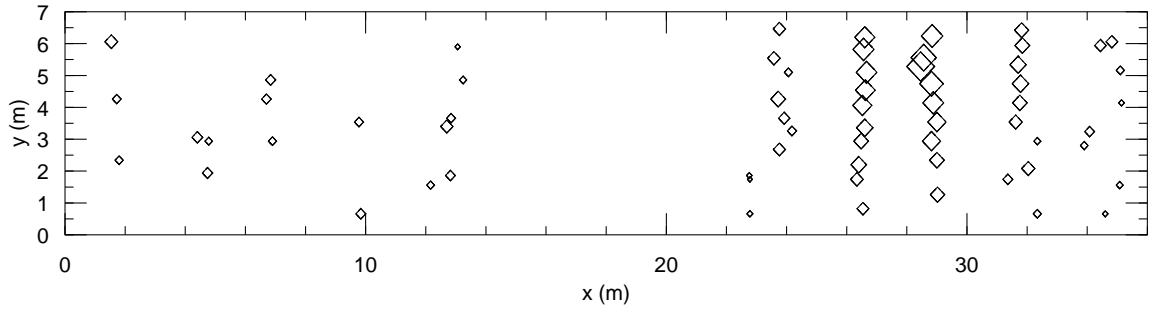,height=4cm} \par  \vspace{1cm}

\caption{ Example of air-shower core  event  observed in the burst detectors.
Rhombi denote the size of events observed in each burst detector and its area
is logarithmically proportional to the burst size.
\label{Fig3}}
\end{center}
\end{figure}

The data set of the burst events analyzed in this paper was obtained during the period 
from October 1996 through  June 1999 \cite{24}. 
First we scan the target maps of all events by the naked eye.  Some events
showing a systematic noise configuration were ruled out during the
first scanning.
An example of the burst detector event is shown in Fig. \ref{Fig3} where the
size of the rhombus is logarithmically proportional to the burst
size. A remarkable lateral distribution in the event pattern is
seen.

   Here, for convenience   we introduce a ^^ ^^ TOP detector'' 
for each burst, which is defined as a detector 
 recording  the highest burst size among all fired burst
detectors.   Furthermore, 
since all the burst detectors are separately set up in 
two sections with a fairly large  distance of 9 m as shown in Fig. \ref{Fig2},
  we call  the section containing the
TOP detector the ^^ ^^ TOP section,'' and the other the
 ^^ ^^ OTHER section'' \cite{24}. 

\begin{figure}
\begin{center}

  \vspace*{1cm}

  \epsfig{file=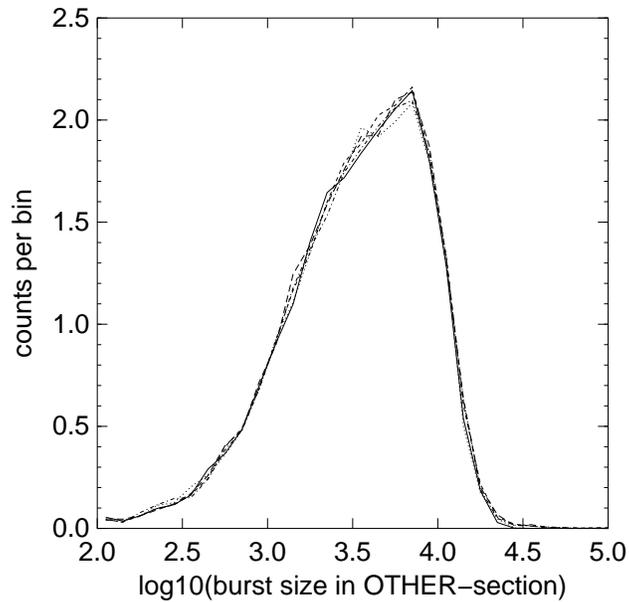,height= 8cm} \par  \vspace{1cm}

\caption{Burst size distribution in the OTHER section. The five curves denote  the different positions of the TOP detector being in the first, second, third, fourth, and fifth
column in the TOP section, respectively. 
\label{Fig4}}
\end{center}
\end{figure}

 We  first examined whether the burst detectors  located
far from the TOP detector still contain signals. For this, we divided the
all burst events into five groups according to
the case that the TOP detector in each event is in the first, second, 
third, fourth,  and fifth  column and then the size distribution observed with
burst detectors in the OTHER section was obtained for each group. If
almost all bursts observed in the OTHER section
are signals, their size distribution must be different from event to
event  because they have different core distances. However 
all these five curves are of the same distribution, as seen in Fig. \ref{Fig4}.

This may strongly suggest that the bursts recorded in the detectors 
far away from the TOP detector, i.e., air-shower core, by more than 10~m 
 are mostly formed by some noises, and its (equivalent) burst size ($N_b$),
which is estimated from the ADC value, is always 
smaller than $3 \times 10^4$  under our experimental conditions as seen 
in Fig. \ref{Fig4}. 
Here, the burst size of $3 \times 10^4$ corresponds to 
a few to 10 TeV for a single gamma ray or a single electron incident
on the surface of the emulsion chamber.
These noises may be mostly induced by an
incomplete ground connection of the detectors to the earth.
Hence, we subtracted the background in the TOP section  assuming that the
same background as in the OTHER section should appear in the
TOP section and they randomly distribute in position. 

After the background subtraction, for a further analysis we made the data set 
by imposing the following conditions on the observed events 
: (1) Size of a TOP detector, $N_b^{top} \geq 10^5$
; (2) size of any non TOP detector, $N_b^{non-top}\geq 10^5$ ; and (3)
 number of fired detectors with $N_b \geq 10^5$, $N_{BD} \geq 1$.  
The total burst size for each burst
event is defined as $\sum N_b$, where the  summation is over  all fired detectors
with $N_b \geq 10^5$. 

5627 events are selected by these criteria, and among them
4274 events are accompanied by  air showers 
with $N_e{\rm (shower~size)} > 10^{4.5}$, which are recorded
by the Tibet-II array.  The time intervals between two
neighboring events are analyzed, and a good exponential distribution is seen,
indicating a good randomness of this data sample. The 
effective running time of this experiment was estimated to be 689.5 days. 
Since the burst detector array was triggered
separately with the Tibet-II array that has a 12\% dead time, 
this value is  taken into account when we calculate the intensity and
 the number of effective events.         

\section{MONTE CARLO SIMULATION}

An extensive Monte Carlo simulation was carried out to simulate the 
cascade developments (air showers) of incident cosmic rays in the atmosphere
and the burst detector responses.
To generate air-shower events in the atmosphere, we used two  simulation codes, 
CORSIKA (+QGSJET interaction model) \cite{25} and COSMOS \cite{26},
both of which are widely used in air-shower experiments. We  also used
an EPICS code \cite{27} to simulate electromagnetic cascade showers in 
the detector. 
In this simulation, the detector  performance, trigger efficiency of detectors, and effective
area are adequately taken into account, based on the experimental data.

\subsection{Primary composition}

    Primary particles we assumed  were   classified
    into seven species as proton (abbreviated to $P$ and mass number=1),
    helium (He, 4), light nuclei ($L$, 8), medium nuclei ($M$ or $CNO$, 14),
    heavy nuclei ($H$, 25), very heavy nuclei ($VH$, 35), and iron group (Fe, 56).
    The absolute flux  of each composition was  fitted  to that obtained
    by direct measurements in the energy region around 1 $\sim$ 10~TeV. The extrapolation
    to higher energies depends on the slopes  of energy spectra and their  bending points.
    As in our previous studies \cite{21,24}, 
    the heavy dominant (HD) and proton dominant (PD)
    models were examined. In HD (PD) the power indices were  assumed to be 2.75 (2.65) for $P$, 
    2.65 (2.65) for He, 2.70 (2.70) for $L$, 2.52 (2.60) for $M$,
    2.60 (2.60) for $H, VH$, and 2.4 (2.60) for Fe, respectively. The bending energy 
    was assumed to be proportional to the charge number and 
    for protons to be 100 TeV in HD, while 2000 TeV for all compositions in PD.
    The fractions of the proton component to the total at 100 and 1000 TeV are
    23 and 11~\% in HD, and 40 and 39~\% in PD, respectively.
     In both cases, the absolute intensity of all particle spectrum 
     was  normalized  so as to be able to reproduce the Tibet and other experimental data
 well \cite{24}. The energy spectra of respective components
 assumed 
in the HD and PD models are summarized in Appendix B.

\subsection{Simulation procedure and simulation data}

   Primary particles at the top of the atmosphere were sampled
   isotropically for the zenith angles within
   $45^{\circ } $.  The minimum sampled energy of primary protons
   was set to  79~TeV and for other nuclei their minimum energies
   are determined so as to keep their contributions from lower
   energies to be less than 1\%. All shower particles were
   followed till 5~GeV by a full Monte Carlo method and then till 1~GeV by the 
   thinning method \cite{25,28}. The shower particles lower than 1 GeV were found to give  minor
  contribution to the burst size since they are absorbed in the lead and iron. 
 The air-shower size of each event
was  obtained using the data calculated by the thinning method.

Each air-shower core which contains all shower particles with energies above 1~GeV was 
thrown on the burst detector array.
Cascade developments of these shower particles in the burst detectors were calculated by 
use of  the analytical
formula which can well fit the full Monte Carlo simulation data obtained by EPICS \cite{24}.
The selection of simulated burst events and their analysis were done
under  the same conditions as used for the experiment.

The events were selected from the simulation data by imposing the same criteria as the 
experiment, and we obtained 
 4 $\times 10^4$  events (9200) for the CORSIKA+HD model (COSMOS+HD). 
Among those selected  events, 50\% (48\%) were induced by protons, 19\% (17\%) by helium,
17\% (15 \%) by $L-CNO$, and 14\% (20 \%) by other heavy nuclei, respectively, while
for CORSIKA+PD,  2 $\times10^4$ events were obtained and the primary ratios 
are 74\%, 16\%, 7.5\%, and 2.5\%,  respectively.
The number of simulated events are 15 times  as many as the experimental data.
It may be worth noting here that the proton-induced events are preferentially 
selected when air showers are tagged by
high-energy cores. That is, even if the primary is heavy-enriched,
almost half of the observed events  selected by the above criteria are 
induced by protons. This is the reason
why we can obtain the primary proton flux from this experiment successfully.

\begin{figure}
\begin{center}

  \epsfig{file=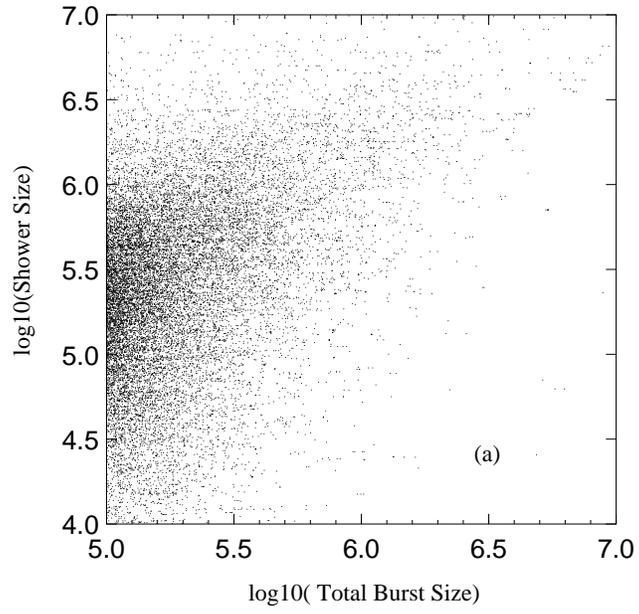,height=8cm} \par

  \vspace*{1cm}

 \epsfig{file=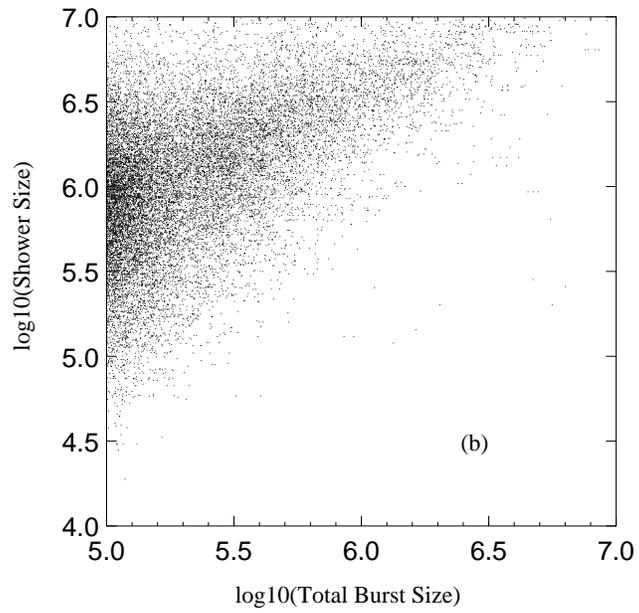,height=8cm} \par   \vspace{1cm}

\caption{ Scatter dots of the total burst size $ \sum N_{b}$ and the shower size $N_{e} $
for the CORSIKA+HD simulation events induced by protons (a) and other nuclei (b).
\label{Fig5}}
\end{center}
\end{figure}

Each event obtained can be  characterized by the following three parameters :
(1) Total burst size, $ \sum N_{b}$ ; (2) total number of fired burst 
detectors, $ N_{BD}$ ; and (3) shower size, $N_{e} $.
Among the three  parameters, $ \sum N_{b}$ and $N_{e} $ are fairly 
sensitive to the primary composition,
as discussed in the previous paper \cite{24}.  The scatter plots
between $ \sum N_{b}$ and $N_{e}$ for the CORSIKA+HD model are  
shown in Fig. \ref{Fig5}. It is seen that  the events with smaller
 $N_{e} $ and larger $ \sum N_{b} $ are mostly
 generated by protons. 
 We use these simulation events in the following analysis.

\section{ARTIFICIAL NEURAL NETWORK ANALYSIS}

As discussed above, the burst events accompanied by an air-shower are 
well characterized by the air-shower size, burst size, and the number of
fired burst detectors. In this experiment, it is also noted  that
 proton-induced events can be characterized by small air-shower size 
and large burst size, while those induced by heavy nuclei have the opposite 
character as their production height is relatively high in the atmosphere because of
shorter mean free path than protons. Based on these facts, 
a simple multivariant analysis was introduced to select proton-induced
events \cite{20}. However, air-shower events are very 
complicated and it is not always obvious what data selection (or cuts)
optimally enhance the signal (proton induced events) over the background. 
Neural networks may be an effective tool since they are ideal for separating patters
into categories (e.g., signal and background). We can train a network
to distinguish between signal and background using many parameters to
describe each event. The network computes a single variable that ranges
from zero to one and if the training is successful the network will
output a number near zero for a signal event and near 1 for
a background event. Hence, a single cut can be made on the network
output which will enhance the signal over the background.

Usually, in a classification problem like the separation of proton-induced events
and others, a set of p events with $k_{max}$ observed variables each,
described by the input vector $\{x^{(p)}\}$ =$(x_1,x_2, ..., x_{k_{max}})$
has to be assigned to output categories $y_i$ using a  set of
classification functions $y_i = F_i(\{x\})$.
For an example, a separation between signal and background events may be based on a
one-dimensional output $y_1$ with the desired value 0 for proton events
and 1 for other events.

\begin{figure}
\begin{center}

  \epsfig{file=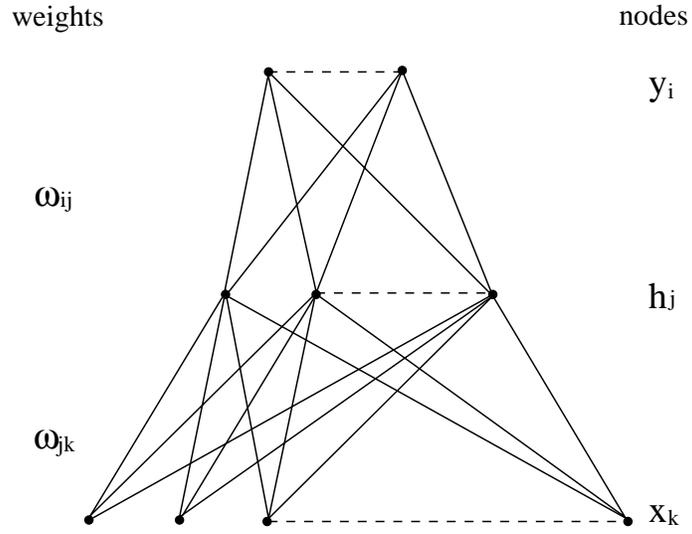,height= 7cm} \par  \vspace{1cm}

\caption{Feed-forward neural network with one hidden layer.
\label{Fig6}}
\end{center}
\end{figure}

For a feed forward
artificial neural network (ANN) with one layer of hidden
units the following form of $F_i$ is often
chosen :

\begin{equation}
     F_i(\{x\}) = g(\sum_j w_{ij}g(\sum_k w_{jk}x_k + \theta_j) + \theta_i),
\end{equation}

\noindent
which corresponds to the architecture of Fig. \ref{Fig6}.
Here, the weights $w_{ij}$ and $w_{jk}$ are the parameters to be fitted
to the data distributions, and  $\theta_i$ and $\theta_j$ are the thresholds
which are generally omitted in the description as they can always be 
treated as weights $\theta_i = w_{i0}$ with $x_0 = 1$.

  $g(x)$ is the nonlinear neuron activation function,
typically of the form (sigmoid function)

\begin{equation}
   g(x) = \frac{1}{2}[1 + \tanh(\frac{x}{T})], 
\end{equation}

\noindent
where $T$ is  a parameter called temperature which is
usually set to 1.

The bottom layer (input) in Fig. \ref{Fig6} corresponds to sensor variables $x_k$
and the top layer to the output features $y_i$ (the classification function $F_i$).
The hidden layer enables nonlinear modeling of the sensor data.
The great success of neural networks is mainly based on the derivation of an
iterative learning algorithm based on gradient descent, the so-called
back-propagation algorithm, and the weights $w_{ij}$ and $w_{jk}$ are 
determined by minimizing an error measure of
fit, e.g., a mean-square error

\begin{equation}
    E = \frac{1}{2} \sum_{p,i} (y_i^{(p)} - t_i^{(p)})^2 
\end{equation}

\noindent
between $y_i$ and the desired feature values $t_i$  with respect
to the weights and $(p)$ is an element of the training data sample.

Changing  $\omega_{ij}$ by gradient descent corresponds to

\begin{equation}
   \Delta \omega_{ij} =  -\eta \delta_i h_j  + \alpha \Delta \omega_{ij}^{old} 
\end{equation}

\noindent
for the hidden to output layers, where $\delta_i$ is given by

\begin{equation}
   \delta_i = (y_i - t_i) g'(\sum_j \omega_{ij}h_j).
\end{equation}

\noindent
Correspondingly, for the input to hidden layers one has

\begin{equation}
   \Delta\omega_{jk} = -\eta\sum_i \omega_{ij}\delta_i g'(\sum_l \omega_{jl}x_l)x_k + \alpha\Delta\omega_{jk}^{old}.
\end{equation}

In Eqs. (4) and (6) $\eta$ is a learning strength parameter which controls
the speed of weight adjustment, and
 so-called momentum terms $\alpha\Delta\omega_{ij}^{old}$ and $\alpha\Delta\omega_{jk}^{old}$ are
included to damp out oscillation. A constant $\alpha$ determines the effect of the
previous weight change. When no momentum terms are used, it takes a long time before
the minimum has been reached with a low learning rate, whereas for high learning rates
the minimum is never reached because of the oscillations.
For a detailed description of the network technique, the back-propagation algorithm
and modifications of the learning rule, see, e.g., \cite{29}

In this analysis, each data set is divided into two parts ;
one that is used for training the network (training data set)
and the other that is used for testing the ability of the network (test data set).  
Then, the whole training data sample is repeatedly presented to the network in a number
of training cycles.
After the network training an independent test data is used to 
check whether the network is able to generalize the classification to the data observed
by our experiment.

 In this work we used a three-layered feed forward network as classifier
of the species of primary particles. That is,
this network contains three parameters as input neurons, ten hidden nodes, and
one output unit and  is abbreviated to a 3:10:1 network. 
Three parameters as input variables are ;
(1) Air shower size $N_e$ ; (2) the number of fired burst detectors  $N_{BD}$,
; and (3) sum of the size of fired burst detector $\sum N_b$.

These are obtained for each event with the detector system consisting of
the Tibet-II array and 100 burst detectors each with an effective
area of 160 cm $\times$ 50 cm.
  The weights in the network were initialized as
uniformly random in the range (0,0.1). The updating of the weights was
done by randomly taking one pattern from
the training set. For overall calculations we used $T=1$ and $\eta = 0.01$.

 Since for the training and test data sample both input $\{x\}$ and correct output
$\{y\}$ have to be known for each event, the adjustment of weights and thresholds depends
on simulated air shower events. For the creation of the training and test showers,
we used the Monte Carlo code ^^ ^^ CORSIKA+ QGSJET'' discussed above. 
The Monte Carlo showers were divided into a training sample and test sample
and  ANN was trained to increase the capability for separating
the proton-induced events from others. The separation power of protons from
others may depend upon the chemical composition of primary particles so
that we trained the ANN using both data samples obtained from the
 HD and PD primary models and checked the  difference between them.

\section{RESULTS AND DISCUSSIONS}

\subsection{Behavior of burst events}

\begin{figure}
\begin{center}

  \epsfig{file=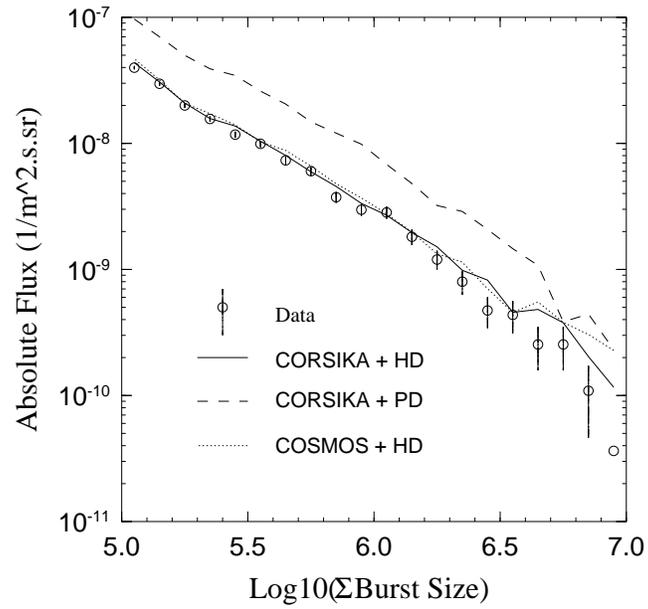,height= 8cm} \par  \vspace{1cm}

\caption{ Burst size spectrum. The open circles, solid line, long-dashed line, and dotted line denote
the experimental data, CORSIKA+HD, CORSIKA+PD, and COSMOS+HD 
simulation events, respectively.
\label{Fig7}}
\end{center}
\end{figure}

First we discuss the behavior of burst events. 
In Fig. \ref{Fig7}, we present the burst size ($\sum N_{b}$) spectrum observed
 in our experiment and
compare it with the simulation results obtained by three different models.

This figure shows that the CORSIKA+HD and COSMOS+HD models are almost consistent with
the experiment.  It is noted that two hadronic interaction models,
QGSJET in CORSIKA and quasiscaling  in COSMOS, can fairly well reproduce many data obtained
by accelerator and cosmic ray experiments.
However, the absolute intensity by the CORSIKA+PD model  gives results about three times 
as high as that by the HD model.
 This difference can be mostly attributed to the difference of the 
 proton flux  in both models since  most selected events are induced 
by protons, in other words, the observed flux of the burst events is very
sensitive to the absolute intensity of primary protons. 

\begin{figure}
\begin{center}

  \epsfig{file=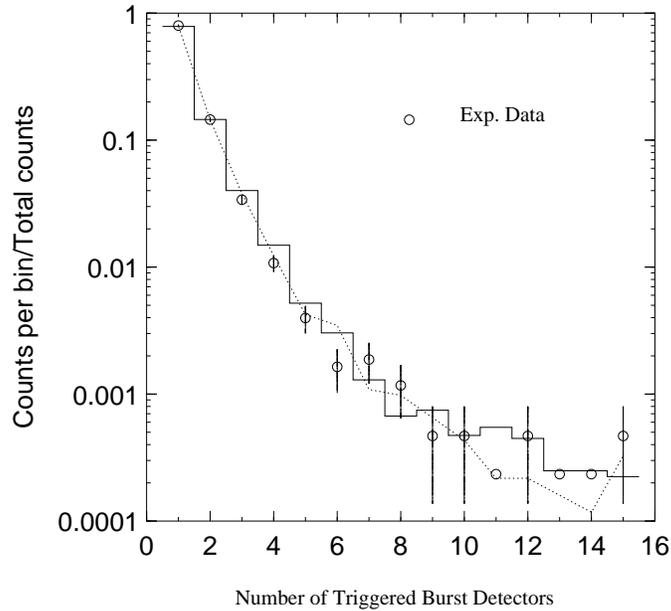,height= 8cm} \par  \vspace{1cm}

\caption{ Number distribution of the fired burst detectors for each event.
 The open
circles, histogram, and dotted line denote the experimental data, CORSIKA+HD
and COSMOS+HD  simulation events, respectively.
\label{Fig8}}
\end{center}
\end{figure}

\begin{figure}
\begin{center}

  \epsfig{file=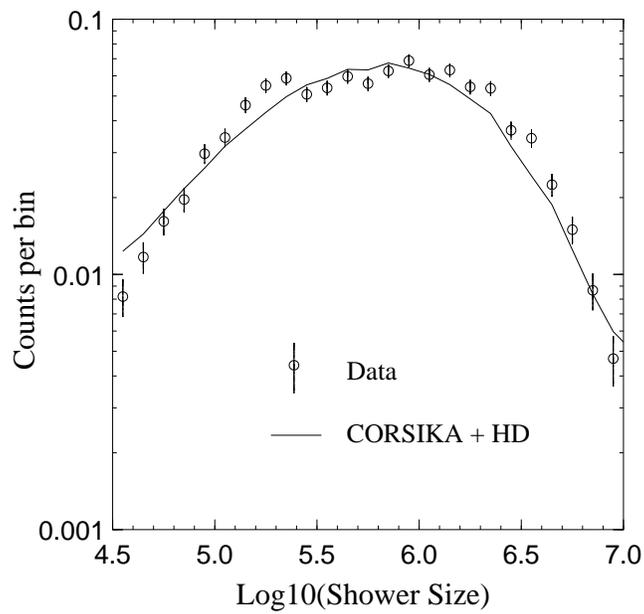,height=8cm} \par  \vspace{1cm}

\caption{ Size distribution of the air showers accompanied by the burst events with
 $\sum N_b > 10^5$.  The open circles
and solid line denote the  experimental data and CORSIKA+HD 
simulation events, respectively.
\label{Fig9}}
\end{center}
\end{figure}

The distribution of the number of fired burst detectors and the air-shower size spectrum
are also shown in Figs. \ref{Fig8} and \ref{Fig9}, respectively, where the experimental
results  are compared with the
simulations obtained by the CORSIKA+HD and COSMOS+HD models. 
From these comparisons,  we can assure  that almost all behavior of the 
burst events observed are compatible with a heavy enriched primary composition 
at energies around the knee. In the previous paper \cite{24}, we also discussed
the detailed features of the burst events whose primary energies are in the knee
energy region, say higher than $10^3$ TeV and reached the same conclusion. 
Based on these results, in the following we try to obtain the primary proton spectrum
from the observed burst events using the ANN discussed above.

\subsection{Selection of proton-induced events with ANN}

\begin{figure}
\begin{center}

  \epsfig{file=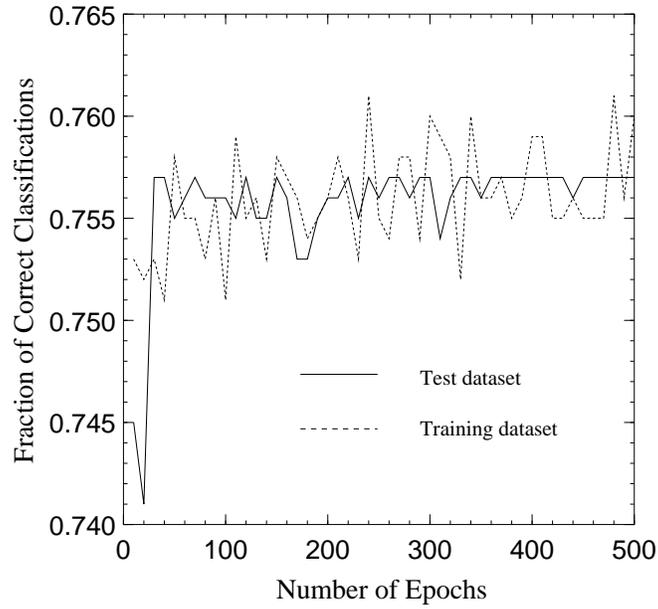,height= 8cm} \par  \vspace{1cm}

\caption{Network performance as a function of the number of 
training epochs, where the dashed and solid lines are
for the training data set and test data set by HD model,
respectively.
\label{Fig10}}
\end{center}
\end{figure}

  We trained and tested the ANN using the simulation events obtained 
from the CORSIKA+HD model, since this model can explain well the behavior 
of the observed burst events as discussed above.  For this, $ 2 \times 10^4$ events by
protons and $2 \times 10^4$  events by other nuclei were used as the training data 
set and the same number of events as the test data set.
The target value for protons was put to 0 and for other nuclei to 1.
A strict middle-point condition was used to measure the classification ability of the 
network, that is, when the ANN output is smaller than 0.5, the event is assigned
as a proton origin, while when the ANN output is larger than 0.5,
the event is considered to be an origin of other nuclei.
The fraction of correct classifications as a function of the number of epochs of the
weight updating is shown in Fig. \ref{Fig10}. The dashed and solid lines are for the 
training and the test data sets, respectively. The learning of the network
becomes very stable after 300 epochs and the change of the weights is small.
It is found that the network is able to correctly select 75.7 \% of the two kinds of
events we input. The wrong classifications are approximately equally
distributed among those two.

As discussed in Sec. III, different primary models give different fractions of the
events produced by each species of primary particles, thus we need to
use different values for cutting the network output in order to
reduce the wrongly classified events to the desired amount.

\begin{figure}
\begin{center}

  \epsfig{file=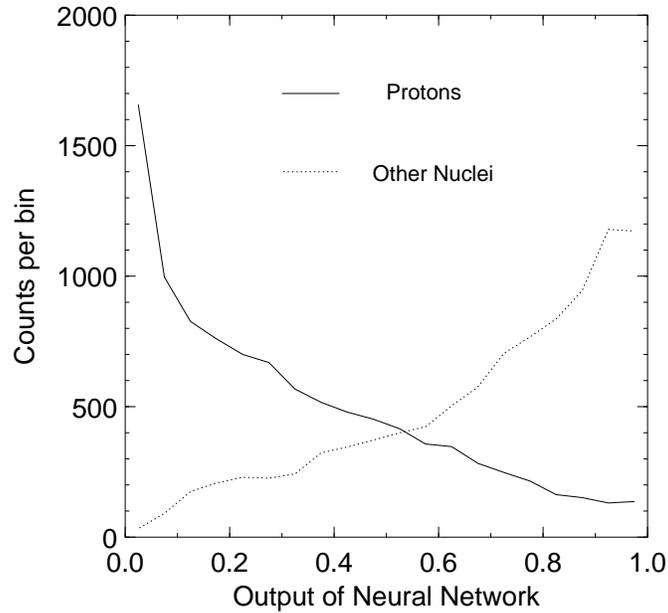,height=8cm} \par  \vspace{1cm}

\caption{ ANN output distribution of the CORSIKA+HD simulation events.
The solid and dotted lines denote the  events induced by  protons and other nuclei,
 respectively.
\label{Fig11}}
\end{center}
\end{figure}

\begin{figure} 
                   \vspace*{4cm}
\begin{center}

  \epsfig{file=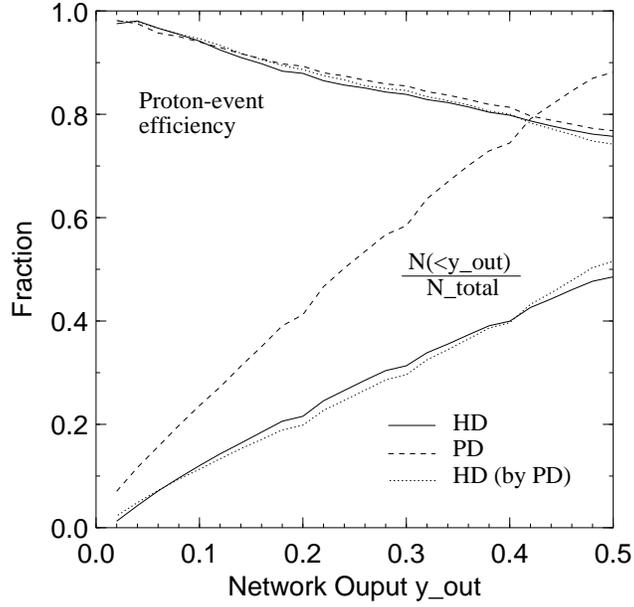,height=8cm} \par  \vspace{1cm}

\caption{ Ratio of the number of selected events with the network out 
smaller than 
$y_{out}$ to the total test events and the selection efficiency of the proton-induced 
events, expressed as a function of the network output $y_{out}$.
Solid line, dashed line, and dotted line are the  cases where the training set is the HD data 
and the test set is the HD data, where the training set is the PD data and the test set
is the PD data, and where  the training set is the PD data and the test set is
the HD data, respectively.
\label{Fig12}}
\end{center}
\end{figure}

 The ANN output distribution of the test events in the case of the HD model is presented in
Fig. \ref{Fig11}. It is seen that the proton-induced events  can be clearly separated 
from others with a  proper cut value  of the ANN output. Shown in Fig. \ref{Fig12} are the
ratio of $N(<y_{out})/N_{total}$ and the selection efficiency of proton events 
as a function of the cut $y_{out}$
in the network output, where $N(<y_{out})$ is the number of events
with the cut $<y_{out}$ and $N_{total}$ is the total number of test events used.
Here we examined three cases : (1) both  training and test data sets
consist of HD events  ; (2) both training and test data sets consist of 
PD events
 ; and (3) training data set consists of PD events while
the test data set consists of HD events. 
As seen in Fig. \ref{Fig12}, it is confirmed that the ANN training is almost  
independent upon
the primary composition and  the selection efficiency of proton-induced events
is about 90 \%  when the cut value of ANN output  $y_{out}$  is set to 0.15.

Using the ANN trained by the CORSIKA + HD events, we selected 820 candidate
events induced by protons out of 4274 observed events.

\subsection{Proton spectrum}

\begin{figure}
\begin{center}

  \epsfig{file=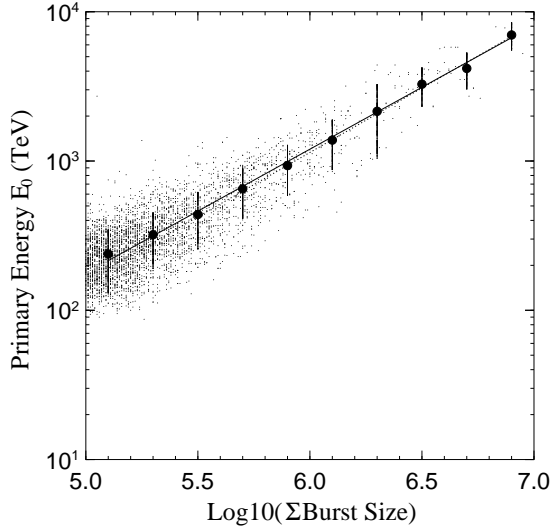,height=7cm} \par  \vspace{1cm}

\caption{ Scatter plots between the primary energy $E_0$ versus the total burst size $\sum N_{b}$ 
 for proton-induced events in the case of the
CORSIKA+HD model. The events are selected by setting 
the ANN output value to 0.15. The solid circles denote the average 
values, and the solid line is a fit by the relation $E_0 = 1200 (\sum N_{b}/10^6)^{0.83}$
(TeV).
\label{Fig13}}
\end{center}
\end{figure}

\begin{figure}
\begin{center}

  \epsfig{file=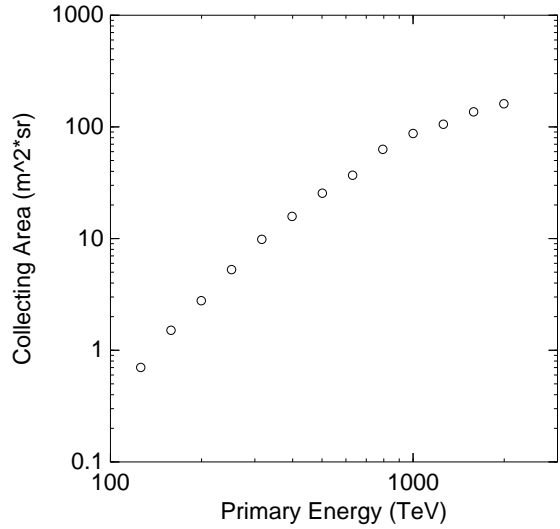,height=7cm} \par  \vspace{1cm}

\caption{ Effective collecting area of the burst array for primary 
protons entering isotropically at the top of atmosphere 
(zenith angle $< 45^\circ$). For the selection criteria
of the burst events, see text.
\label{Fig14}}
\end{center}
\end{figure}

The primary energy of each event can be estimated from a value of $\sum N_{b}$ 
observed with the burst detector array. Shown  in Fig. \ref{Fig13}
is the scatter plots between the burst size $\sum N_{b}$ and the primary energy $E_0$ of 
proton-induced events which were selected from the data set of the CORSIKA+HD events
 by setting the ANN output value to 0.15. 
 A  fairly good correlation  between $E_0$ and $\sum N_{b}$, as seen in this figure,
 enables us to estimate the primary energies  of observed burst events
with small ambiguity. A good correlation is also found between the
air-shower size $N_e$ and the primary energy $E_0$ [ it can be expressed as 
$E_0 \simeq 2.5 (GeV) \times N_e$ for $N_e >10^5$ ], and it is checked
 that both give almost the same
values on the primary energy, while the size estimation
becomes worse for  air-shower events with $N_e < 10^5$ because of small number of detectors
to be used for fitting.
  In the present analysis, then,  we used the burst sizes for 
the estimation of primary energies.  The systematic error  on the primary
energy estimation was evaluated by the Monte Carlo simulation and was
estimated to be about 30 \% at energies around 500 TeV.  Shown in Fig. \ref{Fig14}
is the effective collecting area of the burst array calculated for primary protons 
incident at the top of the atmosphere isotropically within the zenith angle smaller
than $45^\circ$. The burst events satisfying
the selection criteria discussed in Sec. II B and accompanying air showers with 
$N_e >10^{4.5}$ are selected in this calculation.

\begin{figure}
\begin{center}

  \epsfig{file=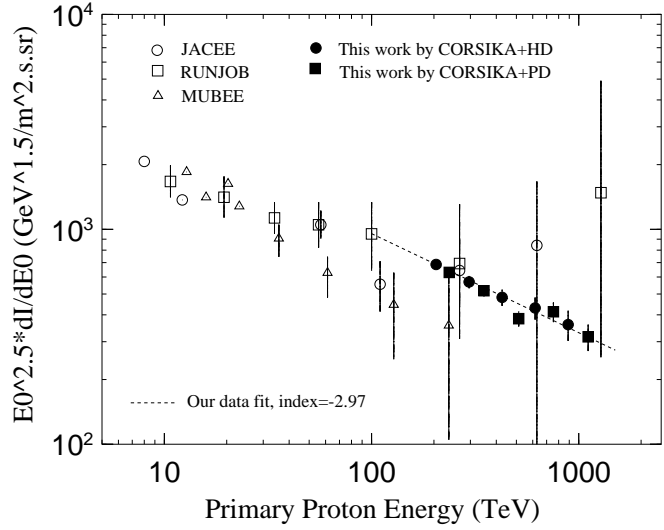,height=7cm} \par  \vspace{1cm}

\caption{ Energy spectrum of primary protons.  The filled circles 
and squares stand  for the experimental results obtained using the ANN's trained 
 by the CORSIKA+HD and CORSIKA+PD events, respectively. 
Our results are compared with other direct measurements by JACEE [10], RUNJOB [30],
 and MUBEE [31]. The dashed line is a best fit to our data.
\label{Fig15}}
\end{center}
\end{figure}

In Fig. \ref{Fig15}, we present the primary proton spectrum obtained from the burst events,
which were selected using the ANN trained by the CORSIKA+HD events.  
To examine whether or not the result depends on the primary composition model used, the following check
was done. For this, first we trained the ANN by using the events obtained from the
CORSIKA+PD model. Then we  selected the proton-induced events from the experimental
data to obtain the proton spectrum.  The primary proton spectrum, thus obtained, 
is also shown in Fig. \ref{Fig15}
to compare with that obtained from the HD composition.
Note that in spite of a big difference between the HD and PD models 
on the  power index and absolute flux of proton component, both results give the
same  spectrum for protons, as seen in Fig. \ref{Fig15}.
Hence,  we may say that the primary proton spectrum obtained from our experiment using
the ANN method is almost independent of the primary composition model
used in the simulation, and it is estimated that the ANN can 
select the proton-induced events from others with an uncertainty of about 10 \%
under our experimental condition.

 The proton spectrum obtained from this experiment  can be represented by
the power-law fit as shown in Fig. \ref{Fig15}.  The power indexes are estimated to be
$-2.97\pm0.06$ and $-2.99\pm0.06$ for the spectra obtained using the ANN trained
by the CORSIKA+HD and CORSIKA+PD events, respectively, 
where errors quoted  are statistical ones. 

It is known that the interpretation of air shower measurements depends 
on the model of the shower development in the atmosphere. The largest
uncertainties may originate from the hadronic interaction which is not well
known at very high energies as well as small momentum transfers. Thus,
using different hadronic interactions may lead to different
predictions for some air-shower observables. No drastic changes, however, 
have been observed on the hadronic interactions at least up to p$\bar{\rm p}$ 
collider energies, corresponding to $\sim$ 1000 TeV in the laboratory system.
Also, it is noted that the air shower size observed at high altitude 
weakly depends on the model, while the difference becomes larger near sea
level \cite{24}.  Furthermore, we examined in the previous paper \cite{24}
that both  CORSIKA (QGSJET) and COSMOS
simulation  codes give almost the same results on the behavior of the burst events observed
with our detector, resulting in that the spectrum obtained here does not
depend on the simulation code we used. 
Consequently, we estimate that the systematic errors on the proton flux are 
smaller than 40 \% in this experiment.

 Direct measurements of the proton spectrum in the energy
region up to about  100 TeV \cite{17,29,30}, while statistics is still scanty,
 may suggest a slightly flat spectrum with the
slope of  -2.5 $-$ -2.7.
When both results are combined, we may say that the proton spectrum changes its
slope at energy around 100 TeV. This may be in favor of shock acceleration
at SNRs  and when we compared this with the all-particle spectrum obtained
by the Tibet air-shower array \cite{17}, the primary composition becomes heavy dominant 
at energies around the knee.

\subsection{On the helium spectrum}

 Our experiment is also sensitive to the helium component.
 In order to estimate the primary  helium spectrum  from our
 experimental data, we adopted the following method.
 Monte Carlo events induced by
 protons and helium nuclei are first gathered as one  group and its ANN target output is
 assigned to be 0, while the events induced by other nuclei 
 belong to another group with the ANN target output 
 being 1.  After training the ANN with the Monte Carlo events, then
 the proton+helium events were selected with a proper cut of the ANN output
 as described in our previous paper \cite{21}. 
 The $\sum N_{b} $ spectrum of the proton+helium events minus that of 
 the proton events should  give  the pure helium spectrum.

\begin{figure}
\begin{center}

  \epsfig{file=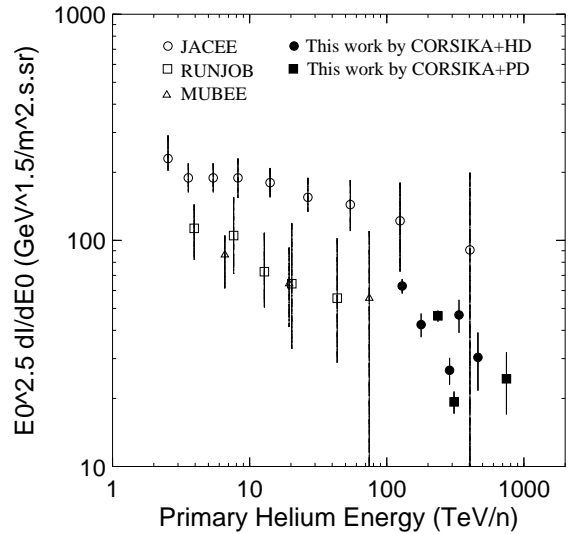,height=7cm} \par  \vspace{1cm}

\caption{ Energy spectrum of primary helium nuclei. For details, see text.
Our results are compared with other direct measurements by JACEE [10], RUNJOB [30],
 and MUBEE [31].
\label{Fig16}}
\end{center}
\end{figure}

 Calculating the effective area for observing the helium-induced 
events  with our burst detectors and also using a 
relation between the  burst size
  $\sum N_{b}$ and the primary helium energy  calculated by the
 CORSIKA+HD model, 
 we obtained the energy spectrum of primary helium nuclei in the energy region 
above about 100~TeV/n, which is shown in Fig. \ref{Fig16}.
 The spectrum obtained based on the CORSIKA+PD
 model is also shown in the same figure to compare with  each other. 
Our data is compatible with
 those extrapolated from the RUNJOB \cite{30} and MUBEE \cite{31} data, and 
the spectrum is not so hard as the JACEE data \cite{17} at high energies.

\section{SUMMARY}

We have been successfully operating a hybrid experiment of burst detector, emulsion chamber
and Tibet-II air-shower array since 1996. Using the data obtained with 
the burst detector array  and the air-shower array and applying a neural
 network
analysis to this data set, we obtained the energy spectrum
of primary protons in the energy range from 200  to 1000~TeV. 
The spectral  index is estimated to be $-2.97\pm0.06$, 
suggesting that the proton spectrum should steepen at energies of 100~TeV
when compared with direct observations done in the lower energy
region.  
 
 We also estimated  the primary helium spectrum at particle energies around 1000~TeV, 
which may have  almost same spectral slope with the proton spectrum, though the statistics is 
 still not enough.

Using  gamma family events, those observed with the  emulsion chamber,
accompanied by air showers, we can estimate the primary proton spectrum
in the energy region from 10$^3$~TeV to $\sim 10^4$~TeV and the result will be
reported in very near future \cite{32}. Then, the Tibet air-shower experiment can
measure the primary proton spectrum in the wide energy range from 200~TeV 
to $\sim 10^4$~TeV and provide vital information necessary for clarifying 
the acceleration mechanism of cosmic ray particles  at very high energies.
     
\acknowledgments

  This work is supported in part by Grants-in-Aid for Scientific
Research and also for International Science Research from the Ministry
of Education, Science, Sports and Culture in Japan and the Committee
of the Natural Science Foundation and the Academy of Sciences in
China. L.K.D., X.W.X., and C.S.Z. thank the Japan Society for the Promotion Sience
for financial support.

\appendix
\section{Performance of the burst detector}

\begin{figure}
\begin{center}

  \epsfig{file=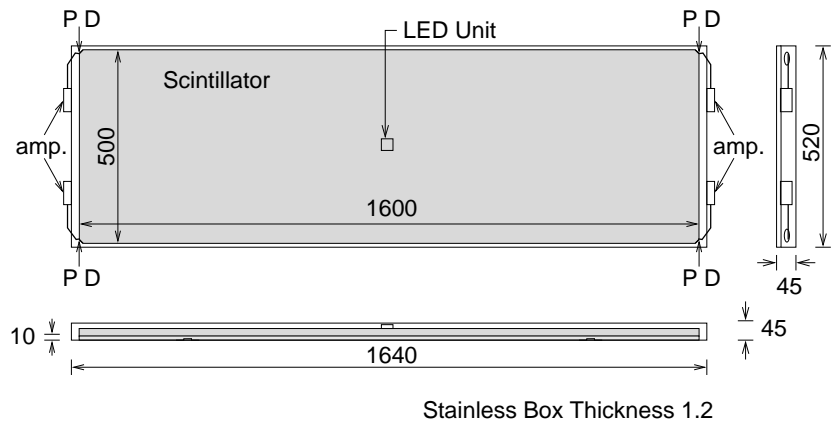,height=6cm} \par \vspace{1cm}

\caption{ Schematic view of the burst detector used in this experiment. Numerals shown in the figure are in units of mm.}
\label{FigA1}
\end{center}
\end{figure}

 Each burst detector contains a plastic scintillator with the size of 
160cm $\times$ 50cm $\times$ 2cm.  
A PIN PD (HPK S2744-03) with an effective area of 2cm $\times$ 1cm was equipped 
at each of four corners of the scintillator, as shown in Fig. \ref{FigA1}.
To detect signals from a PD for burst
particles ranging from 10$^3$ to 10$^7$, a preamplifier with an amplification 
factor of 260 operating in the frequency range from 17kHz to 44 MHz 
(current-current type) was developed. An ADC value from each PD, depending on the size and 
the hit position of a burst (shower) fallen in the
burst detector, can be expressed as  $KN_b(r)$, 
where $r$ is the distance between a PD and the burst position in the 
scintillator, $N_b$ is the burst size,
 and $K$ is  a constant. Using the ADC values from four corners, we can estimate 
the size and hit position for each burst event using a least-squares method.   
In this formula, $f(r)$ denotes the 
attenuation of photons in the scintillator.  
In general $f(r)$ can be expressed as exp(-$r/\lambda$) except at small
 distance $r$ and $\lambda$ takes a value around 350 cm for the present scintillator.
Since the size of the burst detector is smaller than the attenuation length,
 errors of the estimation of burst hit position  become very large.  So we 
first slightly polished 
one face  of each scintillator with rough 
sandpaper (No. 60) to make  photons scatter randomly 
on this face. Then we found  that $f(r)$ can be well approximated 
 as  ${\large r}^{-\alpha}$ and $\alpha$ $\sim$ 1.1 $\sim$ 1.2. This relation was 
confirmed by using a nitrogen gas laser and also cosmic ray muons.  
This dependence on the distance $r$ is sufficient to estimate  
the burst position in the detector.

We also installed  a calibration unit  which consists of four blue light-emitting
diodes (LED's) each having a 
peak wave length of  450 nm. The LED unit is put on the center of each scintillator
 and is illuminated to transmit light
 through the scintillator to each PD at the corner uniformly, and then all the ADC's
 are calibrated at every   10 min for actual run.  This calibration system 
provides information about  a relative change of ADC values, which may cause 
a large  error for the estimation of  burst  hit positions and burst sizes.

We examined the performance of the burst detector using electron beams of 1.0
GeV/c from the KEK-Tanashi Electron Synchrotron. 
The electron beams, ranging from several $\times 10^4$ to
$\sim 3 \times 10^5$ per pulse, were vertically exposed to various positions 
on the surface of the burst detector.

\begin{figure}
\begin{center}

  \epsfig{file=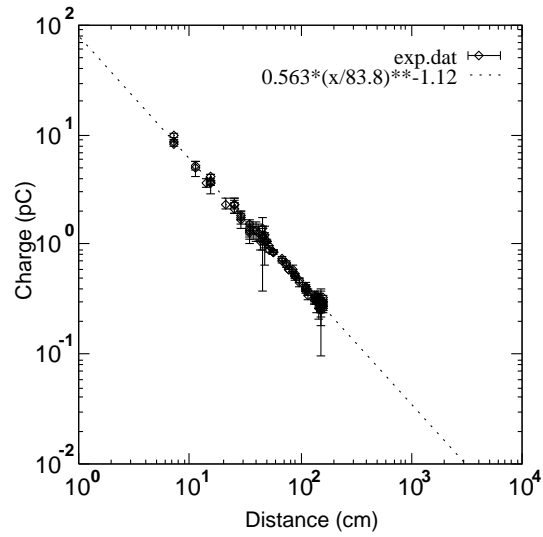,height=7cm} \par \vspace{1cm}

\caption{ Attenuation of photons in the scintillator used for the burst detector, obtained
using electron beams.}
\label{FigA2}
\end{center}
\end{figure}

Figure \ref{FigA2} shows the dependence of the ADC values on the distance $r$, obtained
with the electron beams,
where $r$ is the distance between the beam hit position and  PD. The result can be well 
fitted by a power law of $r$, where the number of incident electrons
 measured by the probe scintillator was normalized to 10$^5$ particles.

\begin{figure}
\begin{center}

  \epsfig{file=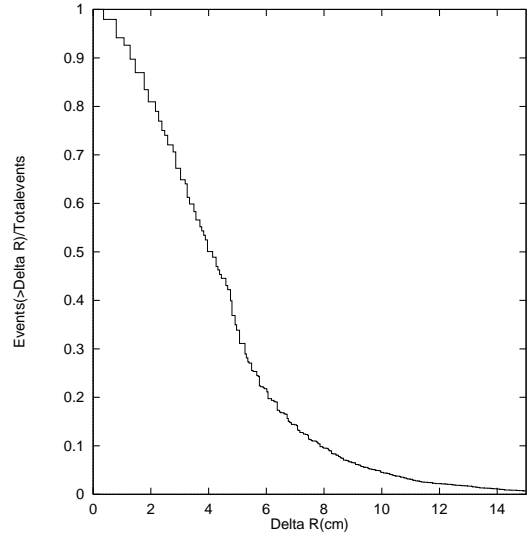,height=7cm} \par vspace{1cm}

\caption{ Distribution (integral) of the difference  between  estimated and irradiated
positions.}
\label{FigA3}
\end{center}
\end{figure}

\begin{figure}
\begin{center}

  \epsfig{file=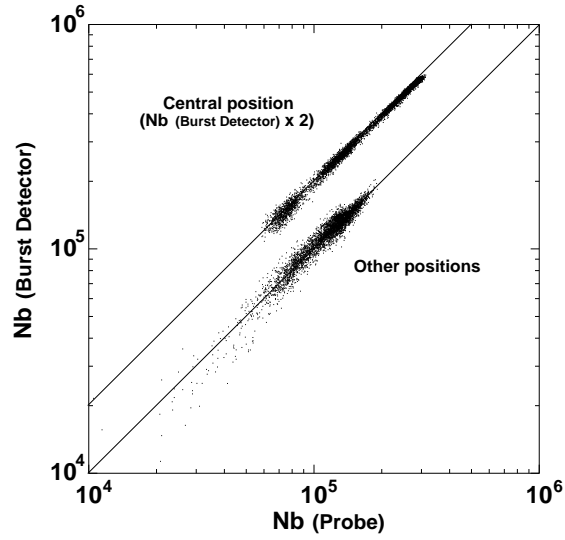,height=7cm} \par \vspace{1cm}

\caption{ Scatter plots of  estimated  and  irradiated number of electrons. 
The number of electrons at various beam positions  on the face of
the detector is normalized to $10^5$ electrons.}
\label{FigA4}
\end{center}
\end{figure}

\begin{figure}
\begin{center}

  \epsfig{file=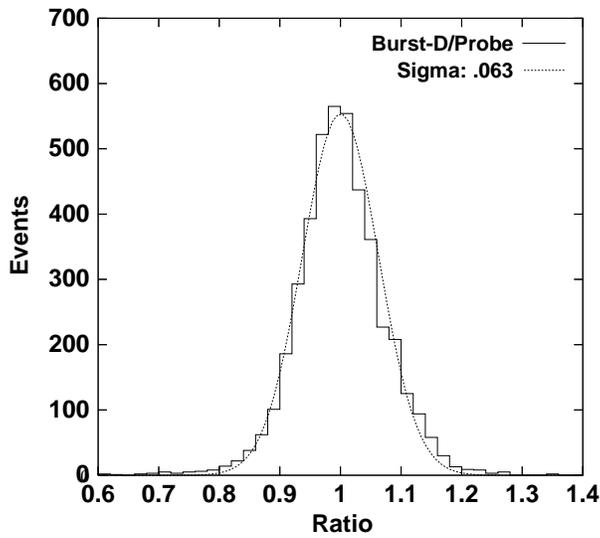,height=7cm} \par \vspace{1cm}

\caption{ Distribution of the ratio of estimated  and irradiated number of electrons
shown in Figure \ref{FigA5}. Dotted line is a Gaussian fit.}
\label{FigA5}
\end{center}
\end{figure}

Using the ADC values from four PD's, the beam positions exposed on the face
of the detector and its intensities (number of electrons) were estimated to
compare with the true ones.
The distribution of the difference between estimated and  actual beam
positions is shown in Fig. \ref{FigA3}.  We  present  scatter plots of the
 estimated number  and  irradiated number of electrons  in Fig. \ref{FigA4},
 and  the  distribution of the ratio between them is  shown in Fig. \ref{FigA5}.
 From these figures, it is concluded that the hit position of 
a burst in each detector can be estimated with an inaccuracy of less 
than 10 cm and
 errors for the size estimation are smaller than  10 \% for the bursts with 
 size $ >10^5$ particles.

\section{Primary cosmic ray composition}

\begin{figure}
\vspace*{-1cm}
\begin{center}
               (a) \par

\vspace*{-3cm}

\epsfig{file=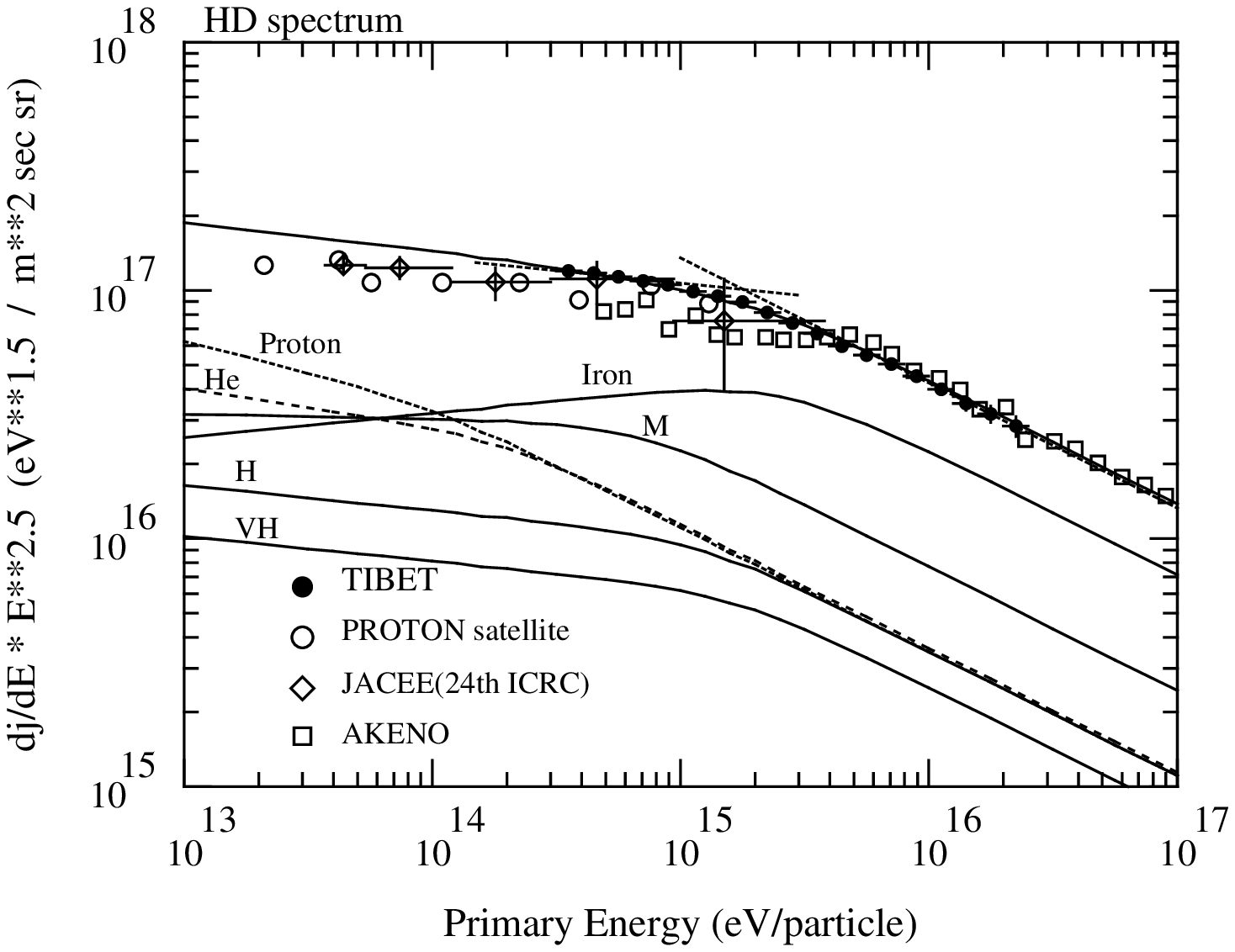,height=11cm} \par 

%\newpage

               (b) \par
\vspace*{-3cm}
\epsfig{file=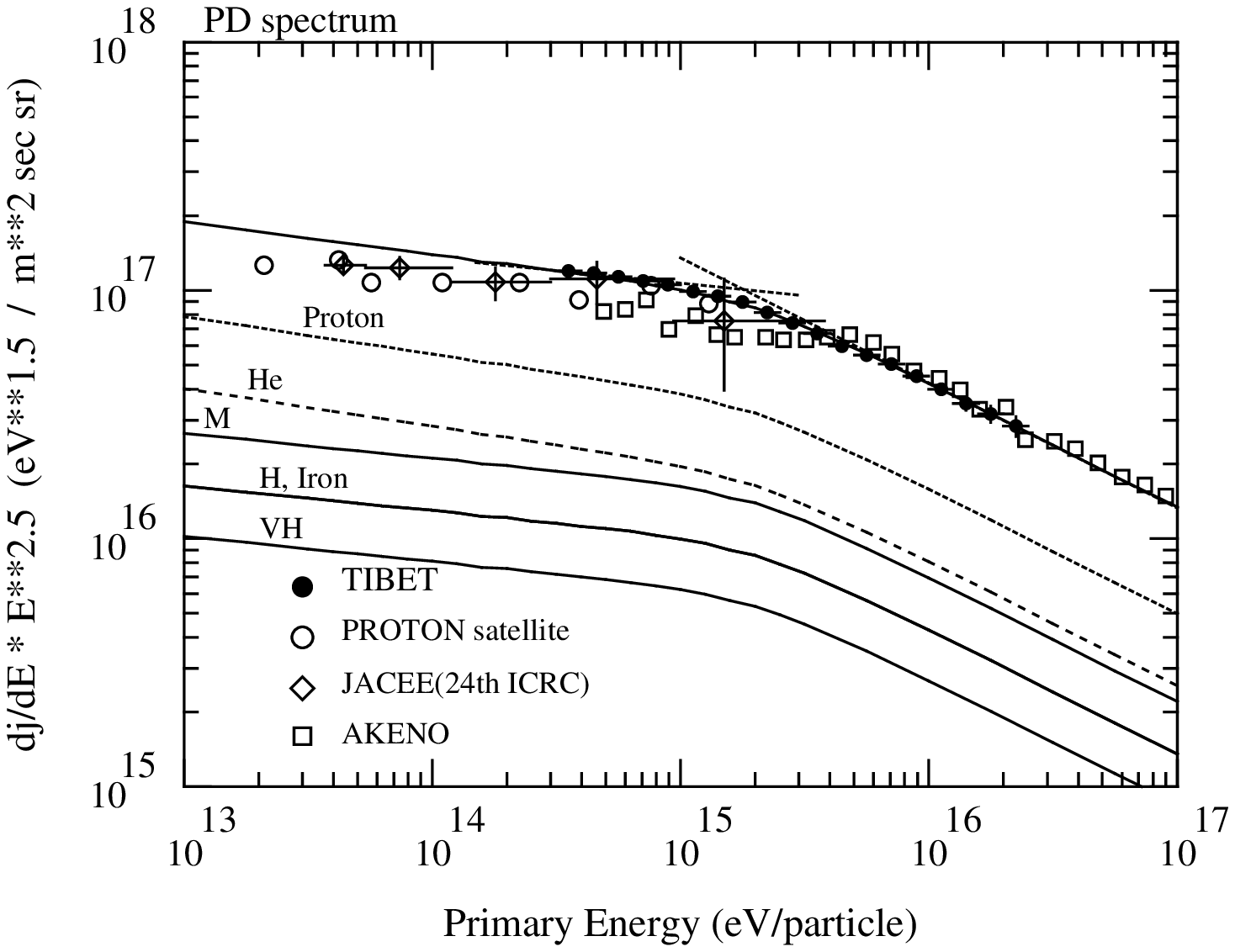,height=11cm} \par

\caption{ Primary cosmic ray composition for (a) the HD model 
and (b) the PD model.  The all-particle spectrum,  which is a sum of
each component, is normalized to the Tibet data.}
\label{FigB}
\end{center}
\end{figure}

The energy spectra of respective components assumed in the heavy dominant (HD)
model and proton dominant (PD) models are shown in Figures \ref{FigB} 
(a) and (b), respectively.  The all-particle spectra obtained
by the experiments : Tibet \cite{17}, PROTON satellite \cite{33},
 JACEE \cite{34}, and AKENO \cite{35} are
plotted in both figures. The all-particle spectrum in each model is normalized to
the Tibet data at energies around the knee.

\newpage

\vspace{2cm}

\vspace*{2cm}

\newpage

\end{document}